\documentstyle[NATO,numreferences]{crckapb}
\begin{opening}
\title{Andreev-Reflection and Point-Contact \protect \\
Spectroscopy of Superconducting Rare Earth \protect \\ 
Transition Metal Borocarbides.}
\author{I.K. YANSON}
\institute{B.Verkin Institute for Low Temperature Physics and Engineering\\
National Academy of Sciences, 61164 Kharkiv, Ukraine}
\end{opening}
\runningtitle{POINT-CONTACT SPECTROSCOPY}

\begin{document}

keywords: rare earth transition metal borocarbides, Andreev-reflection and 
point-contact spectroscopy

\section{Introduction}

The recently discovered \cite{Nagarajan}, superconducting compounds $R$Ni$_2$B$_2$C, 
where $R$ is rare earths (Lu, Tm, Er, Ho and Dy) and Y,  still
remain a hot topic of intensive research. Their crystallographic structure
resembles that of high-T$_c$ materials, albeit being 3D in electronic
properties. The mechanism of Cooper pairing in these compounds is not known
in detail although they are commonly believed to be mediated by the ordinary
electron-phonon interaction, like A15. The boron isotope effect \cite
{Lawrie,Cheon} clearly points that the phonons are involved in
superconductivity . Whether these compounds are $s-$ or $d-$wave
superconductors \cite{Maki} is still under debate. The important sub-class
in these materials is presented by the magnetic compounds ($R=$Tm, Er, Ho, and
Dy), with the antiferromagnetic (AFM) order \cite{Lynn} and weak
ferromagnetism (Er) \cite{wf} coexisting with superconductivity.

The measurements of the quasiparticle density of states (DOS) by tunneling
spectroscopy encounters experimental difficulties, especially these refer to
the magnetic superconductors. To our knowledge, no information is available
about the electron-quasiparticle-interaction (EQI) spectral function
measured by tunneling spectroscopy in the superconducting state, which
unambiguously allows one to determine the mechanism of Cooper pairing.

An investigation of point contacts between a normal metal and a superconductor
can partially solve these problems. The superconducting energy gap and DOS
can be measured by Andreev-reflection spectroscopy \cite{BTK}, and EQI
spectral function can be obtained by inelastic point-contact
spectroscopy (PCS) both in the normal \cite{Yan74} and in the superconducting \cite
{Antalya} states. Comparing the tunneling junction with point contacts, we can see that the
boundary between $R$Ni$_2$B$_2$C-superconductor and vacuum (dielectric) in
tunnel devices constitutes much stronger discontinuity than the interface
between $R$Ni$_2$B$_2$C and a normal metal in point contacts, since in the
latter case there is a Fermi-sea of electrons at both sides around the interface 
with no additional dielectric layer introduced. This discontinuity
may severely influence the measurements making them different from the bulk
properties.

The schematic experimental setup is shown in Fig.1. A small noble metal
(Ag, Cu) rod with sharp edges touches the $(a-b)$-plane edge of a flat $R$Ni$%
_2$B$_2$C single crystal or a sharp edge of a polycrystal with an {\it  a
priory} unknown orientation. Many spots can be tried both on the rod and
along the $R$Ni$_2$B$_2$C edge for investigating the point contacts. The
temperature and magnetic field can be varied, the latter, in case of a single
crystal, is oriented either $\parallel $ or $\perp $ to the $a-b$ plane. The 
$I-V$ characteristics and their first, $R(V)=dV/dI(V)$, and second, $%
dR/dV(V) $, derivatives are recorded by means of a standard lock-in
technique. The latter equals $dR/dV(V)=2\sqrt{2}V_2/V_1^2$, where $V_2$ is
the second harmonic and $V_1$ is the modulation rms voltage.
The diameter of a point contact can be estimated through the Sharvin formula
for a clean orifice $d=27.5/\sqrt{R_0}$ where $d$ is in nm, the nominator is
the average of the noble metal and $R$Ni$_2$B$_2$C \cite{dysp1}, and $%
R_0$ is the zero-bias normal-state resistance of a contact in $\Omega $.

\section{Superconducting energy gap measurements by Andreev-reflection
spectroscopy.}

\subsubsection{BCS-like DOS in paramagnetic and antiferromagnetic state of $%
R $Ni$_2$B$_2$C ($R$=Dy, Tm)}

Recently the temperature dependence of quasiparticle DOS has been measured
in the DyNi$_2$B$_2$C-Ag point contacts \cite{dysp1} by means of
Andreev-reflection spectroscopy. The $R(V)$-curves normalized by the
normal-state $R(V)$ dependences are fitted very well with the
Blonder-Tinkham-Klapwijk (BTK) theory \cite{BTK} with a zero smearing
parameter $\Gamma $ \cite{Dynes}. This evidences that the quasiparticle DOS
is BCS-like in this material. The superconducting energy gap extrapolated to
zero temperature equals (3.63$\pm $0.05) $k_BT_c$ which evidences that on the
average DyNi$_2$B$_2$C is a moderately strong coupling superconductor. The
only disagreement with the BTK theory is that the decrease of the differential
resistance around zero bias in the superconducting state is typically a few
percentage, instead of the order of one-half of a normal resistance \cite{BTK}. 
This can be due either to microscopic inhomogeneity of the material
under the contact where only a small part reveals the unperturbed
superconducting properties, or to the intrinsically decreased Andreev
reflection from an AFM superconductor with ferromagnetically ordered DyC
planes. It should be stressed that DyNi$_2$B$_2$C is the only $R$Ni$_2$B$_2$C
compound which has $T_c<T_N$. Hence, one might think that the molecular
field due to the AFM order averages to zero on the scale of superconducting
coherence length except of very narrow directions almost parallel to the $a-b$
planes.

The situation is different in TmNi$_2$B$_2$C where incommensurate spin
wave ordering occurs at temperatures much below the superconducting
transition $(T_c\gg T_m)$ and the superconductivity develops in the
paramagnetic state. Preliminary results \cite{Yan-Bob1} show that the
overall behavior of $R(V)$-curves also fits the BTK predictions. That means
that this compound also has the BCS form of quasiparticle DOS.

\subsubsection{Fine structure in temperature dependence of superconducting
energy gap in ErNi$_2$B$_2$C near T$_m$.}

The overall temperature dependence of the  Andreev-reflection spectra in ErNi$_2$B%
$_2$C (Fig.2) also follows the BTK model. In this material $T_c>T_m$ and
around $T_m=6.5$ K spin fluctuations depress the superconducting order
parameter which manifested as the lowering of $H_{c2}$ near $T_m$ \cite
{Hc2ER}. Recently, this depression has been measured using the temperature dependence
of the superconducting penetration depth $\lambda (T)$ \cite{Gammel} which
together with superconducting coherence length $\xi (T)$ permits 
determination of the temperature dependence of the thermodynamic critical field $%
H_c=\phi _0/[2\sqrt{2}\pi \xi (T)\lambda (T)]$ proportional to the
superconducting order parameter (energy gap). Direct measurement of the
superconducting energy gap by means of the Andreev reflection and corresponding
BTK fitting (Fig.2) is shown in Fig.3. There is a noticeable dip in $\Delta
(T)$ around $T_m$, but the overall dependence follows the BCS law shown with the
solid line. For this particular contact measured on polycrystalline sample
with unknown orientation, the extrapolated gap is 1.82 mV (2$\Delta _0$%
=4.23 $k_BT_c)$, which is larger than the average value $\Delta
_0/k_BT_c=3.66\pm 0.4$ \cite{RybPB}, probably because of underestimated $%
T_c$. Interestingly, the $\Gamma $-parameter drops around $T_m$ while it
takes relatively large value at other temperatures. The barrier parameter $%
Z $ experiences no change around $T_m$. We emphasize that except the small
anomaly around $T_m$, the quasiparticle density of states in superconducting
state can be approximated by the BCS dependence both in paramagnetic and AFM
states as follows from the BTK fits (Fig.2).

\subsubsection{Two superconducting states of HoB$_2$Ni$_2$C.}

In the Ho-compound, the superconducting transition temperature ($T_c=8.7$ K) is
close to that of the magnetic transition temperature ($T_m\sim 8.5$ K) into the
incommensurate helical state with wave vector {\bf q}$_c^{*}$=0.91{\bf %
c}$^{*}$. Moreover, the incommensurate {\bf a}-axis modulation with 
{\bf q}$_a^{*}$=0.55{\bf a}$^{*}$ also exists over the same
temperature range along with the spiral-AFM transition. Both of these states
collapse near the commensurate AFM transition with {\bf q}$^{*}$={\bf c%
}$^{*}$ and $T_N\approx 5$ K \cite{Lynn}. The Andreev-reflection spectra of HoNi$%
_2$B$_2$C vividly mirror these transformations (Fig.4) \cite
{EuroPhysLet}. Below $T_c$ and down to $T_N$ the $R(V)$-spectra cannot be
fitted by the BTK model. The shallow dip which appears around zero bias has an
order of magnitude larger width than that expected for the given $T_c$. We emphasize
that for $T_c\geq T\geq 6.5$ K the Andreev spectra are completely different
from those in the paramagnetic state of Dy, Tm and Er compounds. Hence, in HoNi$%
_2$B$_2$C we observed the magnetic transition at the same temperature as the
superconducting one, but not at $T_m$=6.5 K as in the single crystals \cite
{Rathnayaka}. It is probable that the uncompensated internal magnetic field along the $a-b$
plane for the spiral structure leads to the gapless behavior of DOS in the
temperature region discussed. Below the AFM transition ($T<T_N=6.6$ K), the
spectra reveal the ordinary BTK behavior. Interestingly, the temperature
dependence of the spectra follows the BCS law albeit with a new $T_c^{*}\approx
6.6$ K. The BTK fits yield a zero-temperature energy gap $\Delta _0=1.04\pm
0.06$ meV which gives $2\Delta _0/k_BT_c^{*}=3.7\pm
0.2$  for lower ''$T_c^{*}$'', i.e. the same as for other $R$Ni$_2$B$_2$C \cite{Yalta}. 
Using the upper $T_c=8.5$ K, one obtains unreasonably low $2\Delta _0/k_BT_c=2.8$.

\section{Point-contact spectroscopy of electron-quasiparticle-interaction
spectral function.}

\subsubsection{Principles of PCS.}

The point-contact spectroscopy (PCS) involves studies of the nonlinearities
of the $I-V$ characteristics of metallic constrictions in the normal state, with
the size $d$ smaller than the inelastic electron mean free path (m.f.p.) 
\cite{KOS,Atlas}. In contrast to a tunneling junction, an ideal point
contact has no barrier. PCS has the advantage that the material is probed
into the depth of the current spreading region, which is of the order of the
constriction size. This size should not be too large in order not to violate the
conditions of the spectroscopic regime of the current flow: $d\leq \min (l_{in},%
\sqrt{l_{in}l_e}),$ where $l_e$ and $l_{in}$ are the elastic and inelastic
electron mean free paths, respectively.

In the ballistic regime ($d\leq l_e,l_{in}$), the second derivative of the  $I-V$
characteristic, $dR/dV(V)$, is proportional to the
electron-quasiparticle-interaction (EQI) spectral function analogous to the
Eliashberg function for the electron-phonon-interaction (EPI):

\[
\frac{d\ln R}{dV}(V)=\frac 43\frac{ed}{\hbar v_F}g_{PC}(\omega )\mid
_{\hbar \omega =eV};\hspace{1.0in}(T\simeq 0) 
\]
Here $g_{PC}(\omega )$ is the EQI or EPI spectral function with kinematic
restrictions imposed by the contact geometry. The factor $4$
(instead of $8$ as in homocontacts) is due to weak EPI coupling and much
greater $v_F$ for the noble metal, as compared to $R$Ni$_2$B$_2$C.
Correspondingly, the noble-metal EPI function is not seen in the spectrum.
For the same reason, in the two-band model of $R$Ni$_2$B$_2$C \cite{Shulga},
only the EQI spectral function for the\thinspace band with the lowest Fermi
velocity and largest EQI is seen in the spectrum.

In the spectroscopic regime no heating of the contact area occurs since the
energy dissipation length, $\Lambda _\varepsilon $, is much larger than the
contact size. However, if the contact size is large compared to $\Lambda
_\varepsilon $, then there is local heating and the temperature rises up
to $T_0\simeq eV/3.63k_B$. The thermal feature in the PC spectra can be
quite large if at a particular temperature $T_{cr}$ a sharp increase in
resistivity occurs like, for example, for the superconducting-normal or
AFM-paramagnetic transitions. The voltage position of such transition on the $I-V$
curve should depend on the input power $V/R^2,$ which means on the contact
resistance.

\subsubsection{Comparison between phonon DOS and EQI spectral function for
superconducting YNi$_2$B$_2$C and non-superconducting LaNi$_2$B$_2$C.}

It is very instructive to compare the phonon density of states, measured by
inelastic neutron scattering, and the EPI (EQI) spectral function, obtained by
PCS, for superconducting and non-superconducting compounds (Fig.5). For
these we choose the nonmagnetic superconducting compound YNi$_2$B$_2$C and
non-superconducting homologue LaNi$_2$B$_2$C \cite{Gompf,EPI-PRL}. The
phonon spectra measured by neutrons contain three groups of peaks:
low-frequency modes (0--30 meV), middle-frequency group (40--60 meV) and
boron high-frequency modes (100--160 meV) \cite{Gompf}. For point-contact
spectra, the high-frequency modes cannot be seen as separate peaks because of 
non-spectroscopic (thermal) regime of current flow where the m.f.p. becomes
shorter than the contact size. For the same reason, the middle-size peaks are
smeared, approaching the thermal regime. On the contrary, the low-frequency
peaks are well resolved and can be compared with the phonon spectra. It should be 
noted that the neutron spectra are taken at room temperature while
the PC spectra at liquid helium temperature. For non-superconducting LaNi$_2$%
B$_2$C the positions of the low-frequency peaks are very close in
both spectra (a) and (b) whereas for YNi$_2$B$_2$C the whole low-frequency
group in PC spectrum (d) is shifted appreciably below the neutron one (c).
This substantial softening of the low-frequency peaks with decreasing 
temperature was thoroughly studied in Ref.\cite{Bullock} by neutron
scattering. A special problem arises with the
low-frequency peak at $\approx 4$ meV, which is not resolved yet. 
In neutron scattering, the emergence
of this peak coincides with the transition to the superconducting state and
it is observed within a narrow solid angle of phonon wave vectors. In PCS it is
seen both in the superconducting and sometimes in the normal state whereas
in the latter case its intensity is greatly diminished by the magnetic field.
Here we remind that in order to measure the PCS spectra the requisite
magnetic field should be applied to suppress superconductivity. One should
also take into account that the lowest-frequency peak is clearly seen in the EPI
spectrum which is the phonon DOS weighted by the averaged EPI matrix element
within a relatively large $\left( \sim 45^{\circ }\right) $ solid angle.

Summarizing the PCS study of EQI function, one can argue that the low frequency
modes are characteristic of the superconducting compounds
while the
behaviour of high energy part of the spectra does not differ appreciably for
superconducting and non-superconducting compounds \cite{YanFNT,EPI-PRL,Yalta}%
.

\subsubsection{Low-frequency phonon and crystal-electric-field excitation
peaks in ErNi$_2$B$_2$C}

In the $R$Ni$_2$B$_2$C compound where magnetic and superconducting orders
coexist (i.e. $R$-ion is magnetic), the new branches of excitations
appear which can interact with an electron. These are the magnons at
temperatures below the characteristic magnetic transition temperatures ($%
T_m,T_N$) and crystal-electric-field (CEF) excitations. Neutron
scattering experiments and magnetization measurements show that the
characteristic energies in the latter case are absent in the range 2-10 meV in
Ho-, and possibly, Dy-compounds \cite{CEF1,CEF2}. In Fig.6 the PC EQI
spectra are shown for two different ErNi$_2$B$_2$C-Ag point contacts. Along
with the second harmonic curves {\bf 1,2}, $V_2(V)$, taken in the magnetic field
needed to destroy superconductivity, the Fig.6 shows (insets)  the Andreev-reflection 
spectra $R(V)$ for the same junctions. These curves serve as a
''passport'' evidencing that there is an intact material under the contact.
Both PC spectra have a large low-frequency peak at about 8-9 meV
confirming that this low-frequency phonon feature is necessary for
observation of the superconducting state. There is an additional structure
at about 6 meV which most probably corresponds to the CEF-excitation
determined in ErNi$_2$B$_2$C by neutrons \cite{CEF-Er1,CEF-Er2}. There is
also a non-identified structure (maximum of $R(V)$) at zero bias which does
not belong to superconductivity.

\subsubsection{Low-energy peaks in HoNi$_2$B$_2$C and DyNi$_2$B$_2$C
compounds.}

The low-energy parts of the EQI spectra of HoNi$_2$B$_2$C and DyNi$_2$B$_2$C are
shown in Fig.7. These compounds possess the same ground state (commensurate
AFM order with magnetic moments aligned ferromagnetically along the [110]
direction on the $R$C planes \cite{Lynn}) though with different $T_N$. Their
spectra are very similar in overall shape while the positions of peaks are
different. The characteristic energies are hardly due only to the
differences between the phonon branches since the masses of the constituent atoms are
nearly the same. Quite definitely, the magnetic excitations should be
involved. This follows from the strong temperature and magnetic field
dependences of the intensity and energy position in HoNi$_2$B$_2$C \cite
{YanFNT,EPI-PRL}. It is difficult to interpret these dependences, since the
magnetic phase diagram in HoNi$_2$B$_2$C is very complicated. Fortunately,
it is simpler in DyNi$_2$B$_2$C. The unexpected temperature dependence of
the lowest-peak for the latter compound is shown in Fig.8 \cite{dysp2}.
At lowering the temperature, the lowest-frequency peak appears at about $%
15 $ K, much above the superconducting order $T_c=6$ K and even the magnetic
order $T_N=10.6$ K. Its height and area grow linearly with $T$ (see the
lower inset), which means that its width remains approximately constant. It
can also be seen in the superconducting state as a strong feature
superimposed on the rapidly changing background due to the superconducting
DOS. The other phonon lines ($eV=10-30$ meV in Fig.8) are smeared due to the
large background in the PC spectra without any noticeable overall change of
intensity with temperature. We are inclined to relate the lowest-lying peak with
magnetic excitations (like the same peak at $eV=3.5$ meV, Fig,7, in HoNi$_2$B%
$_2$C). It should not be the true magnon density of state peak, since it is
observed above $T_N$ in DyNi$_2$B$_2$C, neither it is due to the CEF
excitation, since no peak at this energy is expected. Its origin still
remains a puzzle. Perhaps it may be attributed to the spin fluctuations which,
judging from the neutron measurement, have a large tail at higher
temperatures \cite{Lynn}. In HoNi$_2$B$_2$C, the lowest-frequency peak
mentioned above exists starting at $T=12$ K , which is also above the
magnetic transition temperature $T_m=8.5$ K in this material.

\section{Discussion}

In literature \cite{Carter,Vaglio} one can encounter a statement that at least the
non-magnetic compounds $R$Ni$_2$B$_2$C (Y, Lu) are similar to A15
superconductors in that they have enhanced electron density of state at the
Fermi energy and their Cooper-pairing mechanism involves phonons \cite
{Lawrie,Cheon}. Still, despite several attempts, no phonon structure was
observed above the energy gap in the tunneling characteristics \cite
{tun1,tun2,tun3,tun4}. In the tunneling characteristics of A15, 
a distinct phonon structure is observed with a
proper value of the electron-phonon parameter $\lambda $ \cite{Wolf}. For
magnetic $R$Ni$_2$B$_2$C, the situation is even worse, since the
superconducting energy gap is not safely measured \cite{tun1}, to
say nothing about the above the gap spectroscopic structure. This situation is
disturbing and similar to that in high-T$_c$ superconductors where no structure
is seen above the gap, yet \cite{highTc1,highTc2}.

Contrary to tunneling, the PC spectra show strong electron-quasiparticles
interactions both with phonons and magnetic excitations. In principle, they
can be studied both in normal and superconducting \cite{Antalya} states.
Interpretation of these spectra is not straightforward, since several
electron-quasiparticles interactions can interfere with each other. The
magnetic structure can hinder the electron-phonon interaction responsible
for Cooper pairing \cite{Morozov,Amici}. On the other hand, the
superconducting transition may lead to a change in magnetic order \cite
{Norgaard} since it affects the RKKY interaction. Even in the non-magnetic
compounds, the proximity of the superconducting energy gap to 
nonadiabatic softening of phonon modes may lead to drastic changes in the
phonon spectrum and electron-phonon interaction. Only the systematic study
of PC spectra, both in non-magnetic and magnetic compounds in different
magnetic fields and at various temperatures, can shed light on the peculiarities of Cooper
pairing in these compounds. In this respect, the use of high quality single
crystals is invaluable, since the properties of these compounds are often 
very anisotropic.

\pagebreak

\begin{figure}
\caption[]{Schematic presentation of a sample holder.
}
\end{figure}

\begin{figure}
\caption[]{Differential resistance of polycrystalline ErNi$_2$B$_2$C-Ag point contact
at different temperatures: $T=$ 4.2, 4.6, 5.0, 5.6, 6.2, 6.8, 7.4, 8.2, and
9.0 K from bottom to top. The experimental curves (dots) are shifted
vertically  for clarity and normalized for each curve separately. The BTK
fits are shown with thin straight lines. $H=0.$ Resistance in the normal
state: $R_N=16.5$ $\Omega $.
}
\end{figure}

\begin{figure}
\caption[]{(a) Temperature dependence of the superconducting energy gap in ErNi$_2$B$_2$%
C (solid square dots) determined by BTK fitting shown in Fig.2. The dashed
vertical straight line points to the magnetic ordering temperature $T_m$.
(b) and (c) are smearing parameter $\Gamma $ and barrier strength $Z$ for
the same point contact, respectively. Note the dip of the gap at 
$T_m$.
}
\end{figure}

\begin{figure}
\caption[]{Temperature dependence of Andreev-reflection spectra for HoNi$_2$B$_2$C. The
curves represented by dots are shifted vertically for clarity after
subtraction of a polynomial fit of the 8.6 K data \cite{EuroPhysLet}. Near
each curve the temperature is indicated. $T_c$ and $T_c^{*}$ stand for
superconducting transition temperature and the ''BCS-like'' transition
temperature, respectively. BTK fits are shown as thin solid lines. $%
R_N=2.65$ $\Omega $. The Y-scale is shown by a vertical bar of 0.1 $\Omega 
$. Note that the BCS-like Andreev-reflection spectra appear only below $%
T_c^{*}$.
}
\end{figure}

\begin{figure}
\caption[]{Comparison of phonon densities of states measured by incoherent scattering of
neutrons at room temperatures \cite{Gompf} (curves (a) and (c) ) and
electron-quasiparticle spectral functions obtained by point-contact
spectroscopy at liquid helium temperatures \cite{EPI-PRL} for
non-superconducting LaNi$_2$B$_2$C and superconducting YNi$_2$B$_2$C,
respectively. Note the substantial softening of low-frequency phonons in
YNi$_2$B$_2$C due to lowering  temperature\cite{Bullock}.
}
\end{figure}

\begin{figure}
\caption[]{The second harmonic signal for two different ErNi$_2$B$_2$C-Ag point contacts
proportional to the EQI spectral function. Resistance of junctions 
{\bf 1} and {\bf 2} are 8.25 and 14.4 $\Omega $, modulation voltages
0.975 and 1.94 mV, and magnetic field 1.95 and 2.6 T, respectively.
Correspondingly, for each junction their Andreev-reflection spectra are
presented in the inserts. Note the low-frequency peaks at energies below
10 meV and the feature at $eV\simeq $6 meV which is tentatively ascribed to
crystal-field excitation.
}
\end{figure}

\begin{figure}
\caption[]{Comparison between the PC spectra of HoNi$_2$B$_2$C and DyNi$_2$B$_2$C. The
parameters are $R_N$=2.3 and 27 $\Omega $, $T=4.2$ and 1.8 K, and $H=0.5$
and 0.65 T, for HoNi$_2$B$_2$C and DyNi$_2$B$_2$C, respectively
}
\end{figure}

\begin{figure}
\caption[]{The temperature dependence of the EQI PC spectra of DyNi$_2$B$_2$C with $%
R_N=4.8$ $\Omega $, $V_1=0.76$ mV. The temperatures are 6, 7, 9, 11, and 15 K
for different curves, respectively. Either the area or the height defined as is shown 
in the upper inset for $T$=6 K are used for the intensity of the lowest-lying peak.
The lower inset displays the temperature dependence of
these parameters (squares - area, dots - heights) normalized at $T=7$ K.
Three more pairs of dots are added which correspond to the other contact (not shown) whose
characteristic is presented in Ref.\cite{dysp2}. Note the onset temperature $%
\approx 15$ K which is substantially higher than $T_N=10.5$ K and $T_c=6\,$
K. $H=0.$
}
\end{figure}

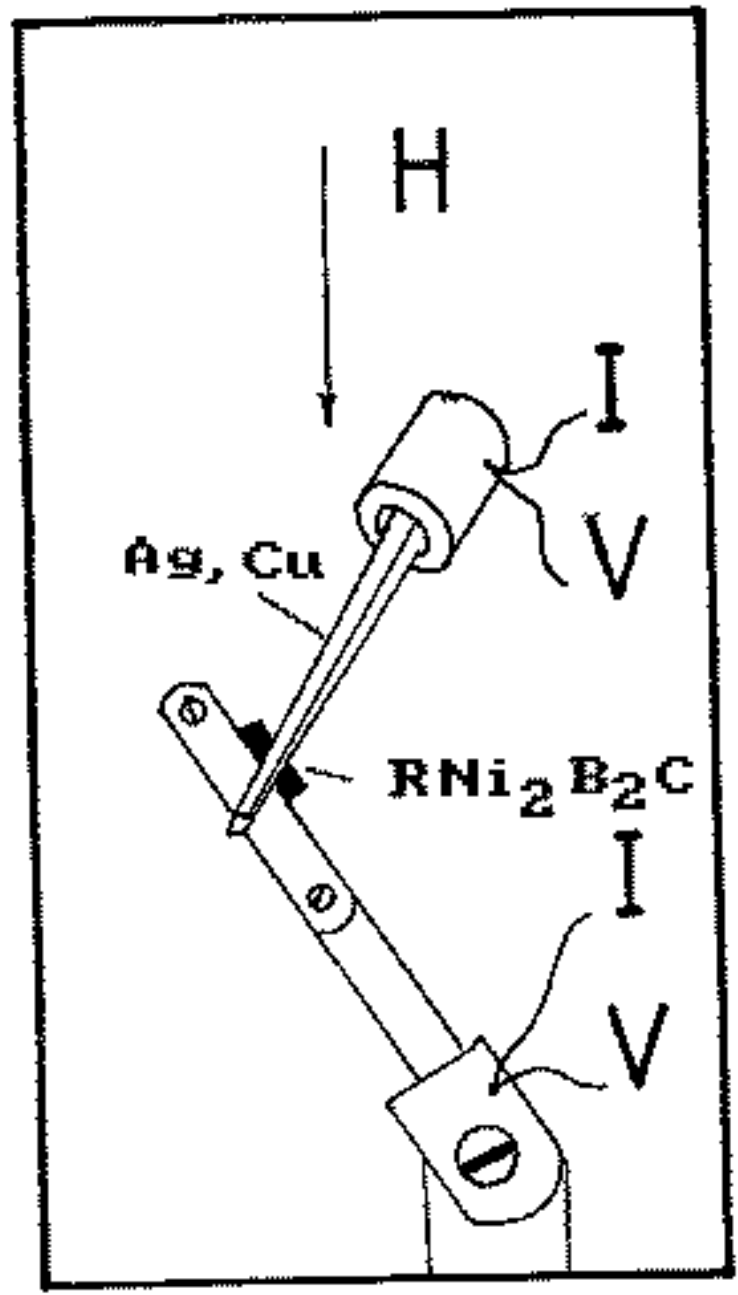
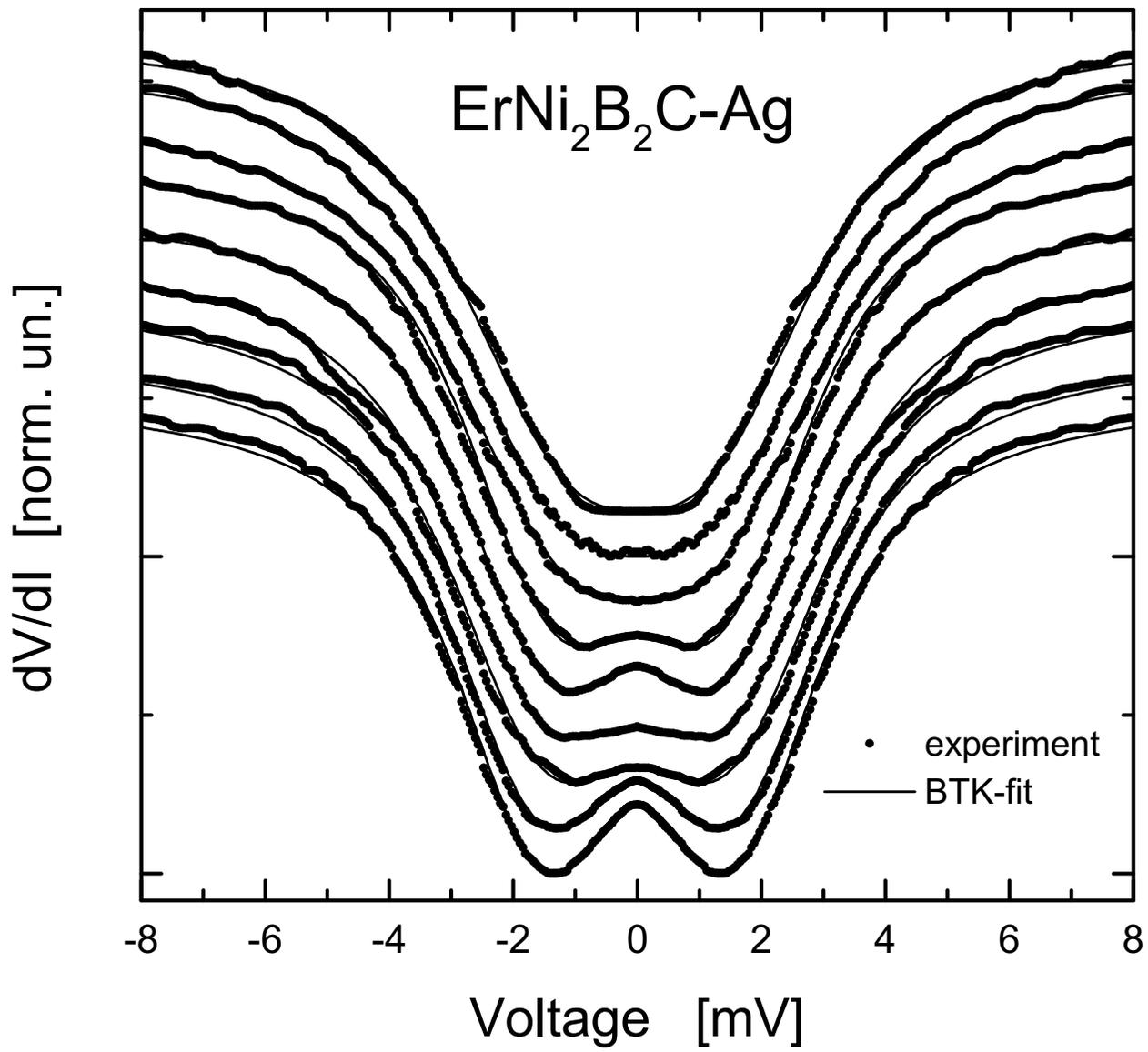
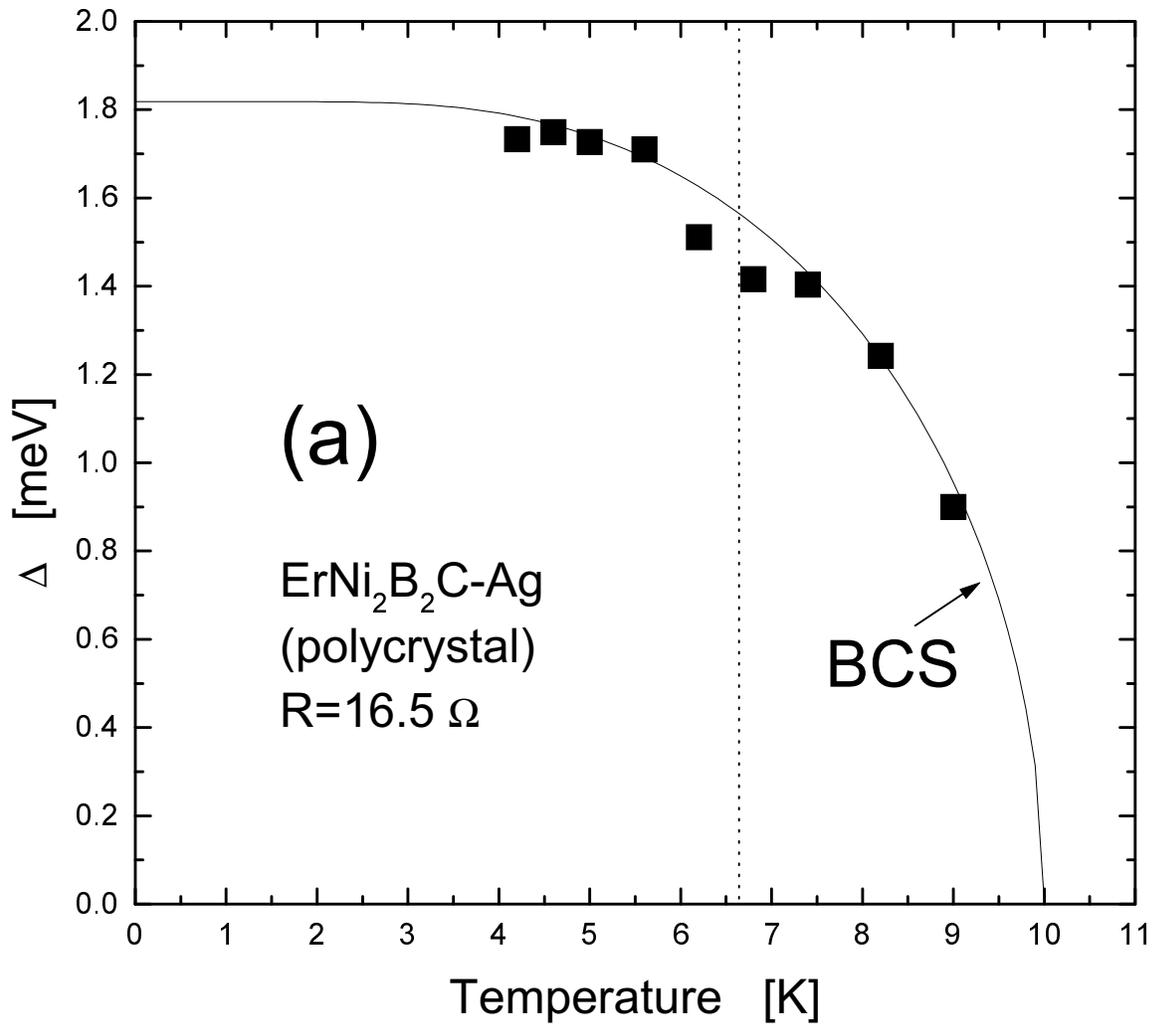
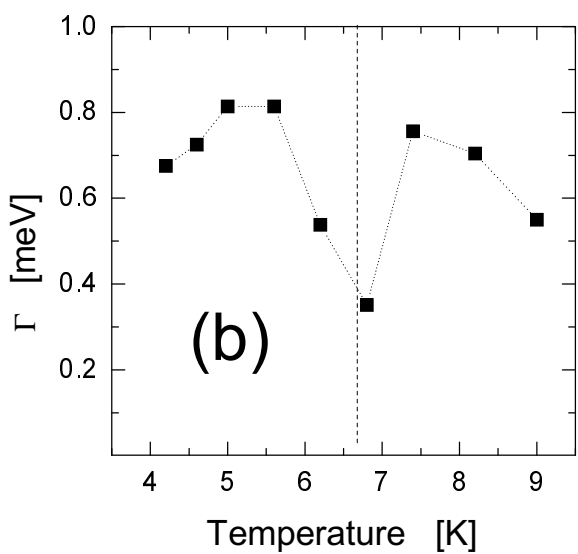
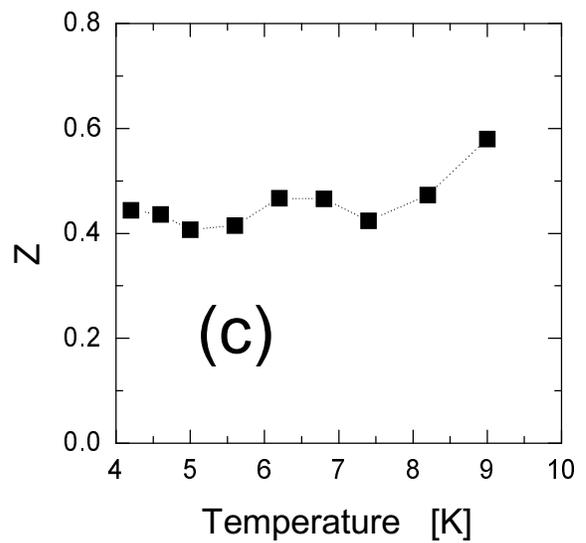
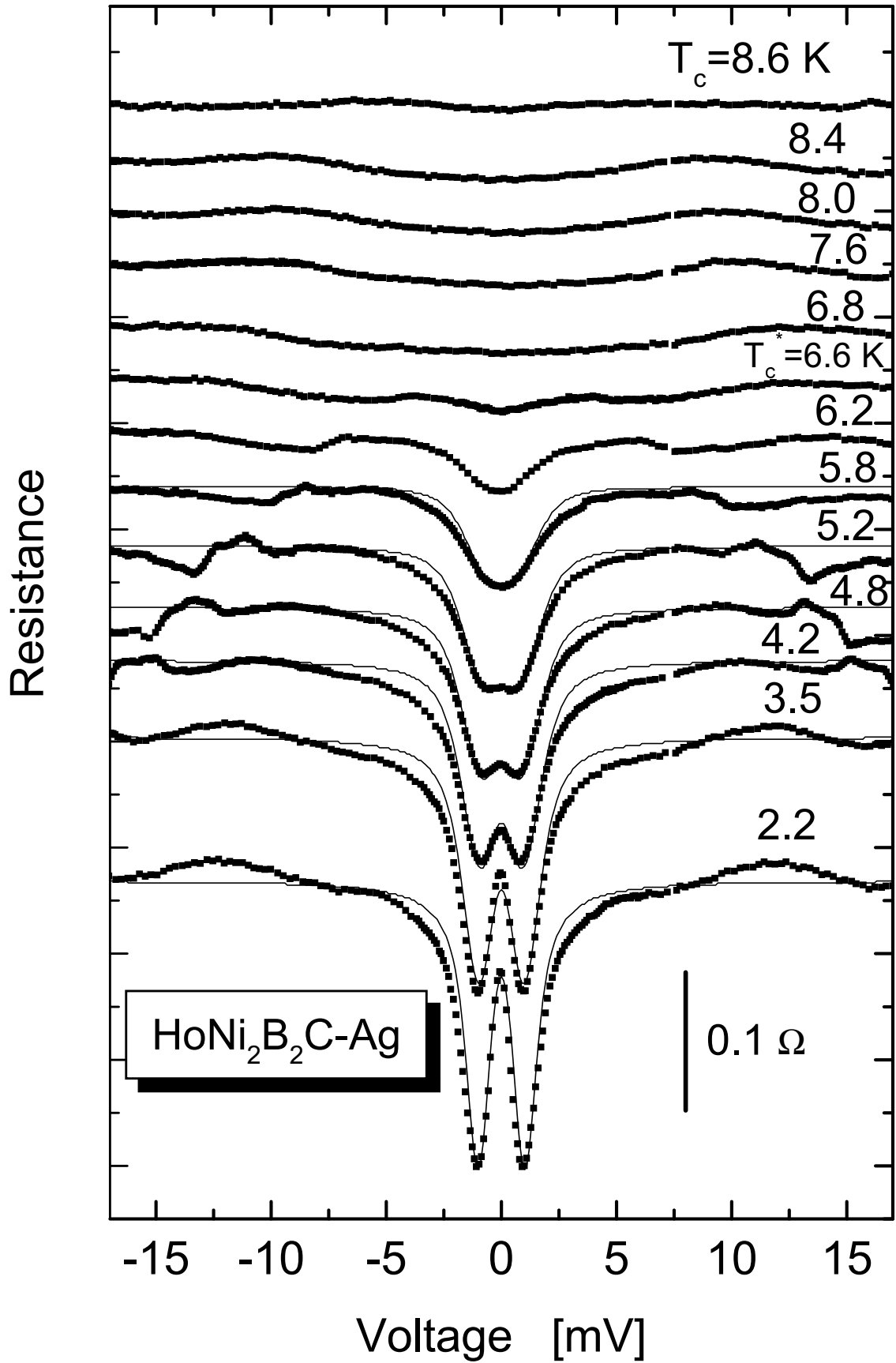
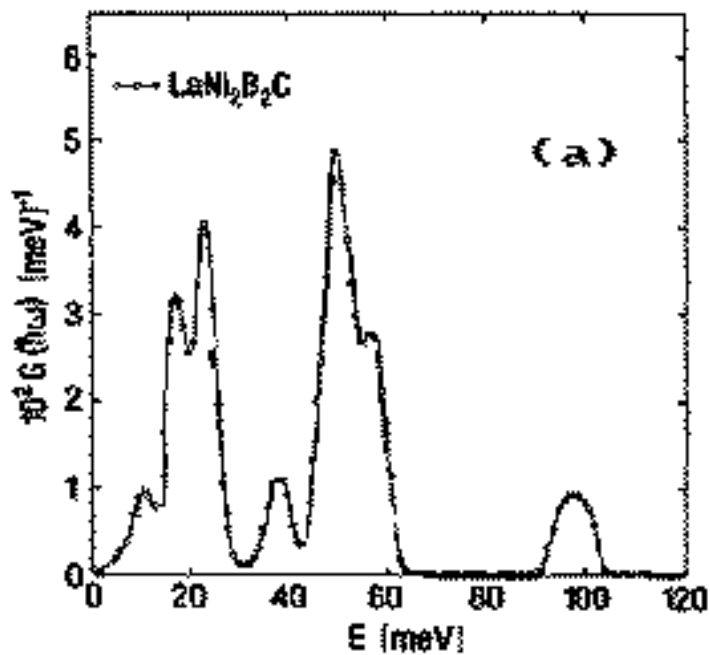
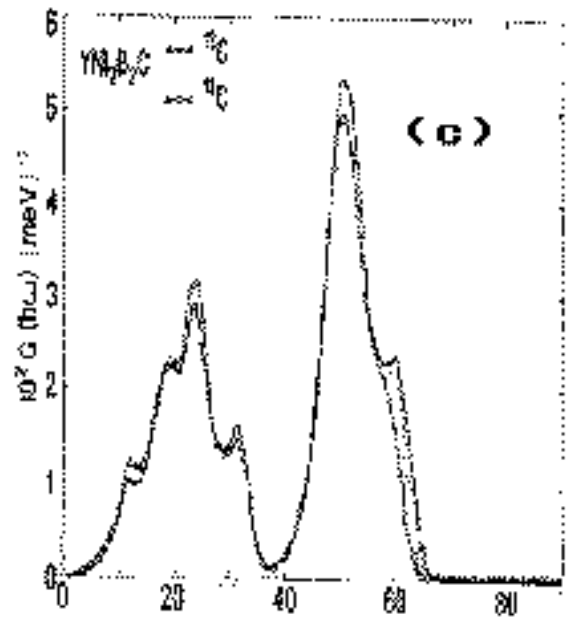
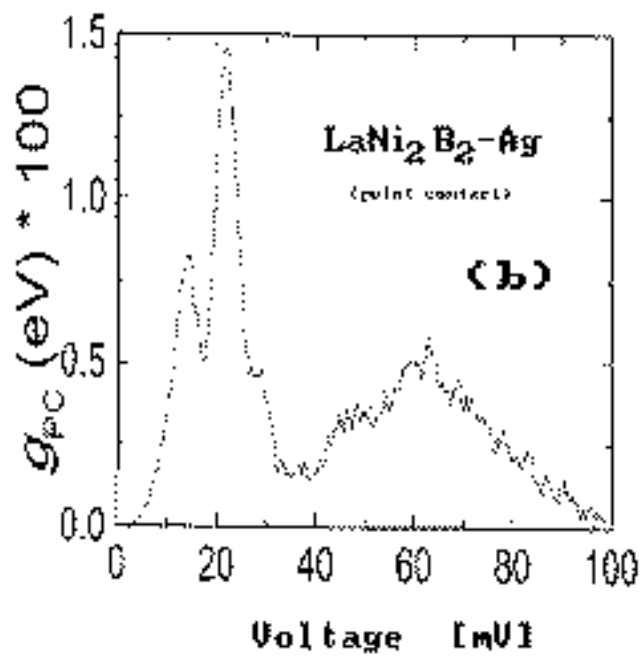
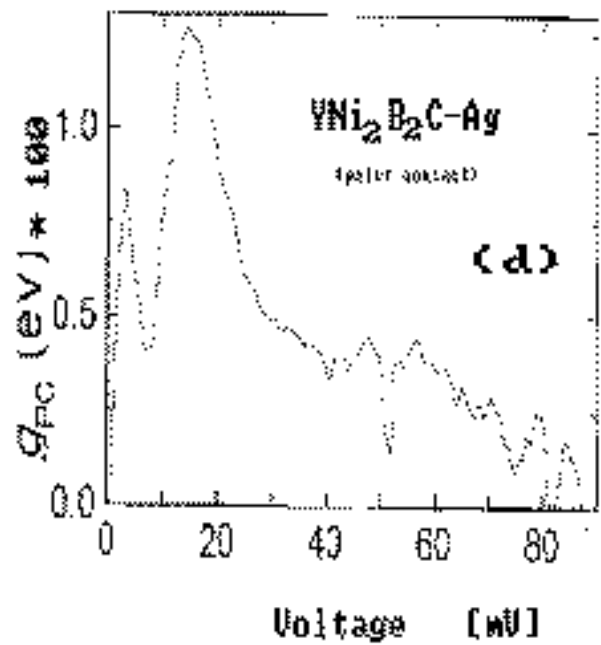
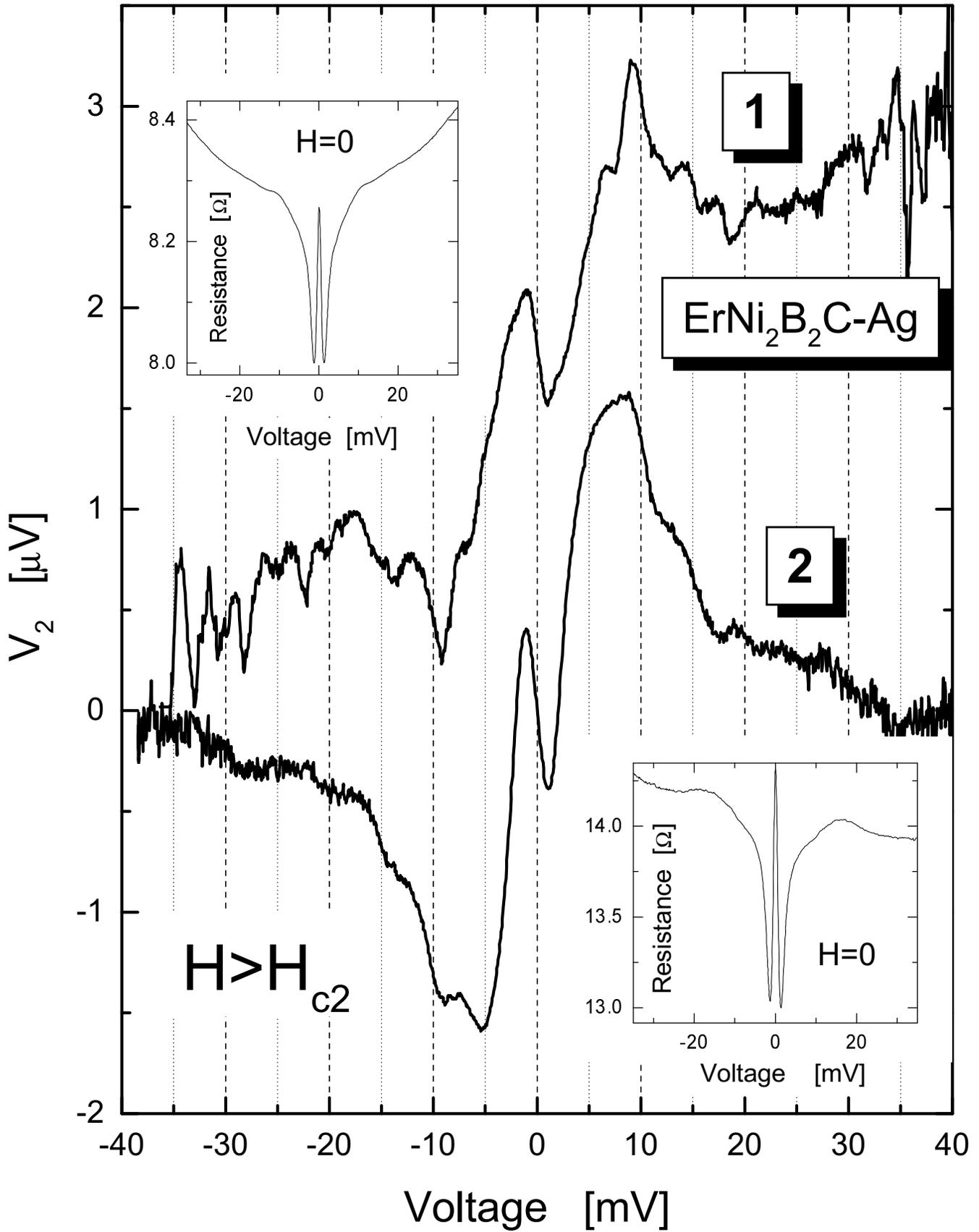
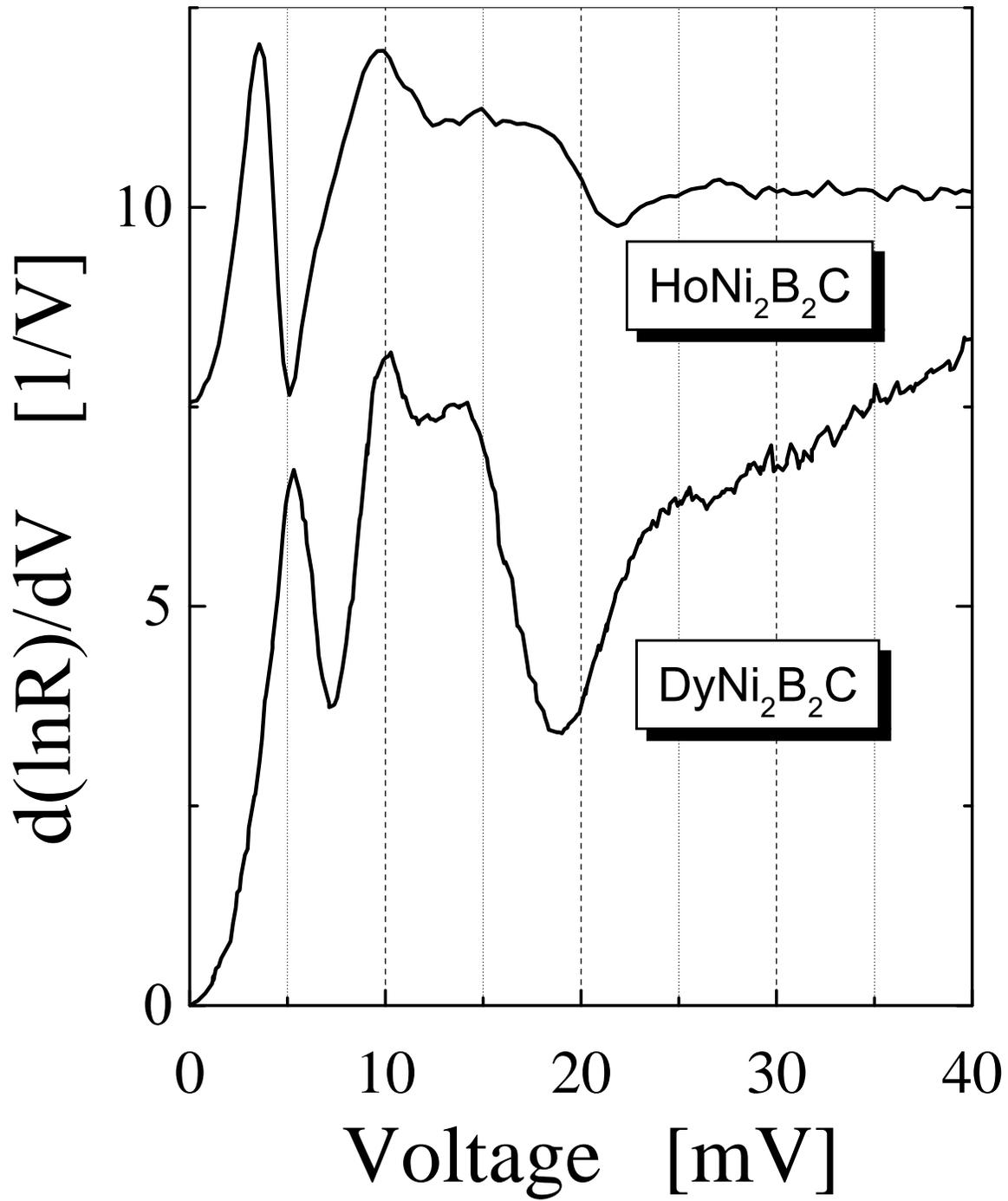
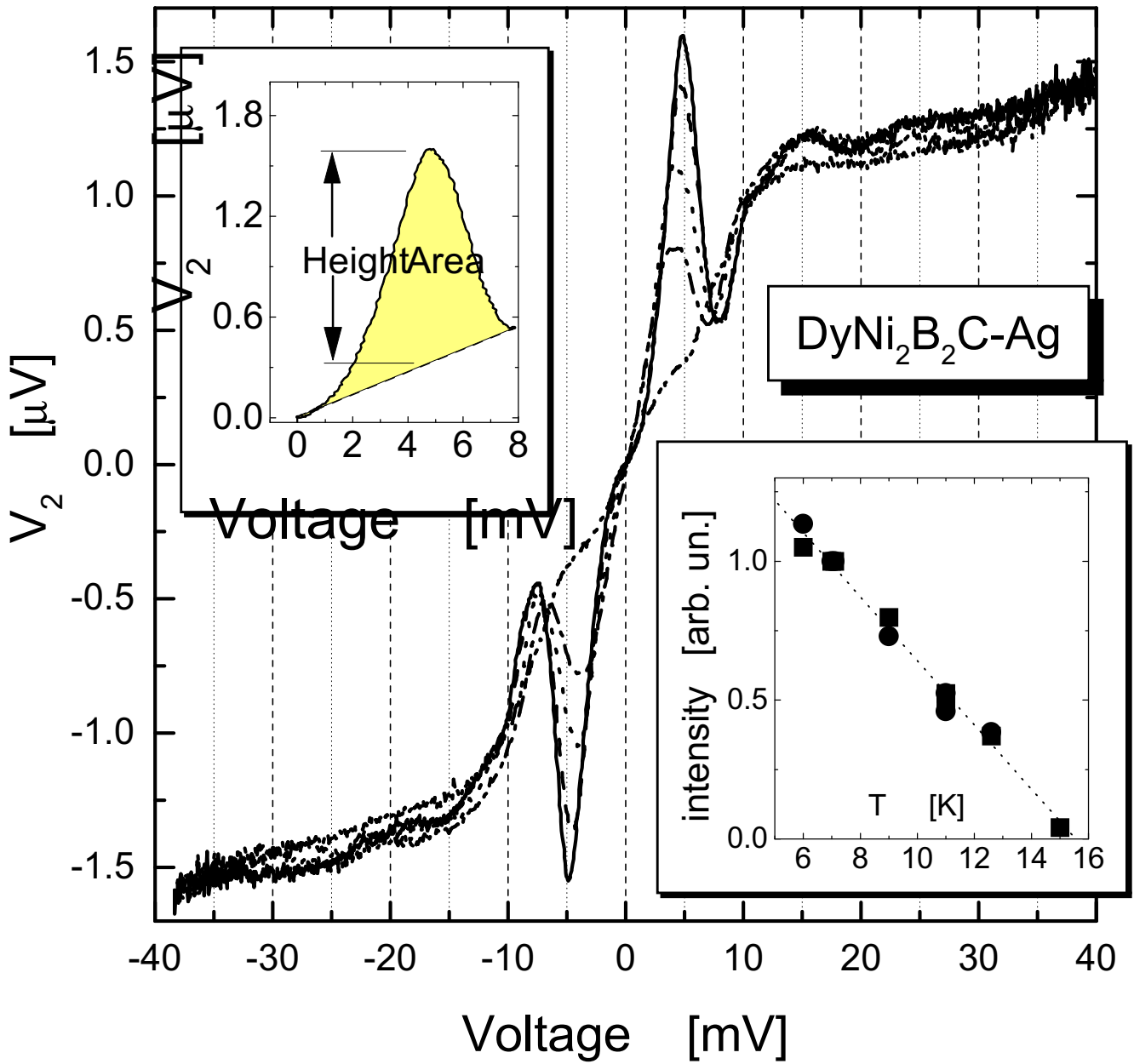
\end{document}